\documentclass[aps,prl,showpacs,floats,twocolumn,amsfonts,amssymb,unsortedaddress,floatfix]{revtex4}
\input epsf
\usepackage{bm}
\usepackage{amsmath}
\usepackage{graphicx}
\usepackage{xcolor}
\usepackage{ulem}

\begin{document}

\addtolength{\topmargin}{10pt}

\def\Bbb{\mathbb}

\title{Attractive Inverse Square Potential, $U(1)$ Gauge,  and  Winding Transitions} 

\author{Cristiano~Nisoli$^1$ and A. R. Bishop$^2$}
\affiliation{$^1$Theoretical Division, $^{1,2}$CNLS, and $^2$Directorate for Science, Technology, and Engineering 
\\
Los Alamos National Laboratory, Los Alamos NM 87545 USA}
\email[]{cristiano.nisoli@gmail.com}

\begin{abstract}
The inverse square potential  arises in a variety of different quantum phenomena, yet notoriously it must be handled with care: it suffers from  pathologies rooted in the mathematical foundations of quantum mechanics. We show that its recently studied conformality-breaking corresponds to an infinitely smooth winding-unwinding topological  transition for the {\it classical} statistical mechanics of a one-dimensional system: this describes the the tangling/untangling  of floppy polymers under a biasing torque. When the ratio between torque and temperature exceeds a critical value the polymer undergoes tangled oscillations, with an extensive winding number. At lower torque or higher temperature the winding number per unit length is zero. Approaching criticality,  the correlation length of the order parameter---the extensive winding number---follows a Kosterlitz-Thouless type law. The model is described by the Wilson line of a (0+1) $U(1)$ gauge theory, and applies  to the tangling/untangling  of floppy polymers and to the winding/diffusing kinetics in diffusion-convection-reactions.
\end{abstract}

\pacs{03.65.Vf, 64.70.Nd, 87.15.Zg, 11.15.-q, 36.20.Ey}

\maketitle 
The quantum mechanics of the  Inverse Square Potential (ISP)~\cite{Morse53,Case50}  is an old problem that has attracted much recent  attention~\cite{Gupta, Marinari, Moroz10, Camblong00, Kaplan09,Essin, Nishida07, Martinez}. It is relevant to phenomena as diverse as the Efimov effect for  short range interacting bosons~\cite{Efimov70, Efimov73} (recently confirmed experimentally~\cite{Kraemer06}), the interaction between an electron and a polar neutral molecule~\cite{Levy, Camblong01}, the near-horizon problem for certain black holes~\cite{Claus, Camblong03},   the anti-de Sitter/conformal field theory correspondence~\cite{Witten98}, and nanoscale  optical devices~\cite{Denschlag97}. In statistical mechanics, the inverse square potential represents the borderline case for  phase transition for the long-ranged 1-D Ising model~\cite{Thouless69, Dyson71, Luijten08, 40Years}.

While the mathematics of  ISP is well understood~\cite{Morse53,Case50},  its practical use  remains often problematic~\cite{Gupta, Marinari, Moroz10, Camblong00, Kaplan09,Essin, Nishida07, Martinez}. The quantum mechanics of its conformally-invariant hamiltonian  is well posed for repulsive or weakly attractive couplings, yet  it is not self-adjoint for strong attractions~\cite{Simon, Simander, Narnhofer}, leading to unphysical pathologies typical of singular potentials~\cite{Frank}. Most relevantly,  its  bound spectrum is a continuum and unlimited from below~\cite{Morse53, Case50}. The problem is often rendered physical  by a short-distance cutoff, when possible, or by other renormalizations~\cite{Gupta, Camblong00}: in all cases conformality is lost either by regularization or by a renormalization anomaly.
 
 When regularized by finite cutoff, the potential produces an infinite but discrete and limited spectrum of bound states and negative energies, with a defined ground state. At  the crossover between strong and weak attraction the bound states disappear in the same fashion as the inverse correlation length in the Kosterlitz-Thouless  (KT) transition~\cite{Kosterlitz73}, a feature  of loss of  conformality~\cite{Kaplan09}. This is suggestive of an associated (topological) transition. 

Here we  show that  the  problem can indeed be related to an infinitely smooth topological transition for  a one-dimensional system which is biased to wind around the pole of a non-simply connected  space. The order parameter, as we shall see, is then the average winding number, which approaches zero infinitely smoothly at transition while its correlation length follows a KT law.  
The transition is thus between the winding and non-winding functional sub-manifolds of the hamiltonian, which can be made topologically distinct by boundary conditions. The removal of these boundary constraints at transition  corresponds to the extension of the gauge space for a $(0+1)$ $U(1)$ symmetry, whose Wilson  line describes our system.  

For physical definitiveness, the model can describe a tangling/untangling  phase transition for floppy polymers~\cite{Edwards,DeGennes, Flory} under torque: current single molecule manipulation techniques~\cite{Neuman} could  test it experimentally. 
Not surprisingly, it is also associated with a kinetic transition for a diffusion-convection-reaction in  a screw dislocation.  

Consider first the  one dimensional ``Hamiltonian'' for the attractive ISP
\begin{equation}
\hat H=-\frac{1}{2 \chi}\left(\frac{d^2}{d\rho^2}+\frac{\gamma^2}{\rho^2}\right),
\label{Hs}
\end{equation}
which can be made self-adjoint for  $|\gamma|>1/2$   by a short distance cutoff $\rho\ge\bar \rho$ such that (\ref{Hs}) acts on smooth functions of $\left[\bar \rho,\infty\right]$ with a Dirichlet (zero)
boundary condition at $\bar{\rho}$~\cite{Gupta}. 
We  associate to it  the partition function (or density operator)
\begin{equation}
Z_{\gamma}(\rho_l, \rho_0)=\int_{_{\rho(0)=\rho_0, \rho>\bar \rho}}^{^{\rho(l)=\rho_l}}\!\!\!\!\!\! \!\!\!\!\!\! \!\!\!\!\!\!  [d\rho]\exp{\left[-\int_0^l\left(\frac{\chi}{2} {\dot \rho}^2-\frac{\gamma^2}{2\chi\rho^2}\right)ds\right]}, 
\label{Zeta2}
\end{equation}
such that (with the above  regularization  for  $\hat H$)~\cite{Kleinert}:
\begin{equation}
Z_{\gamma}(\rho_l, \rho_0)\propto\langle \rho_l| e^{-l \hat H}|\rho_0\rangle.
\label{trace}
\end{equation}
(Equipartition factors in $\chi$, irrelevant to the transition, are neglected in the following.)
We then introduce  
\begin{equation}
\langle \omega \rangle=l^{-1}\partial_{\gamma} \ln Z_{\gamma}=\langle k^{-1} \rangle \gamma,
\label{Omega}
\end{equation}
as an order parameter. The second equality in (\ref{Omega}) defines the average winding compliance $\langle k^{-1} \rangle$,   the reciprocal of a generalized rigidity. Equation  (\ref{Omega}) can  be rewritten as  ${\langle k^{-1} \rangle=2 l^{-1}\partial_{\gamma^2} \ln Z}$, and then from  (\ref{Zeta2}) we find
\begin{equation}
\langle k^{-1} \rangle=\langle {1}/{\chi \rho^2} \rangle.
\label{kgen}
\end{equation}
As we will see below, as $\gamma \to (1/2)^+$ the order parameter $\langle \omega \rangle$ disappears  in the thermodynamic limit ($l\to\infty$) and thus the generalized rigidity diverges.

The physical meaning of the above treatment becomes clearer by performing  a Hubbard--Stratonovich transformation on (\ref{Zeta2}) in the auxiliary variable $\omega$, which yields 
\begin{equation}
Z_{\gamma}(\rho_l, \rho_0)\propto\int_{_{\rho(0)=\rho_0, \rho>\bar \rho}}^{^{\rho(l)=\rho_l}}\!\!\!\!\!\! \!\!\!\!\!\! \!\!\!\!\!\!   [\rho{\kern 0.1em}d\omega{\kern 0.1em}d\rho]{\kern 0.2em}\exp{\left(-\beta\int_0^l{\cal H}ds\right)}
\label{Zeta}
\end{equation}
for the energy per unit length ${\cal H}$, given by
 \begin{equation}
 \beta{\cal H}=\chi \left(\dot \rho ^2+\rho^2 \omega^2\right)/2-\gamma \omega.
 \label{H}
 \end{equation}
(Here $\beta=1/T$, the Boltzmann constant is taken $k_B=1$). Note that, as a consequence of the transformation, there are no boundary conditions on $\omega$. 

The $\rho$ in the functional measure $[\rho{\kern 0.1em}d\omega{\kern 0.1em}d\rho]{\kern 0.2em}$ is due to the inverse Gaussian integration over $\omega$ and  the functional measure is thus reminiscent of  a surface element in polar coordinates. Indeed
 (\ref{Zeta}), (\ref{H}) control the statistical mechanics of a field $\psi(s)=\rho(s) \exp i \int^s\omega(s') ds'$, which describes trajectories in a punctured (because  $\rho>\bar \rho$) complex plane. Then, from (\ref{H}),  the order parameter $\langle \omega \rangle$ in (\ref{Omega}) is simply the average linear density of winding  number (up to a factor $2\pi$)  of these trajectories around the pole: the quantum-mechanics of  ISP with ultraviolet cutoff  is thus turned into the statistical mechanics of a one-dimensional object winding around the pole of a non-simply-connected plane. 

For a system described by this model (see the conclusion for more realizations)
consider  a floppy  polymer---a random chain~\cite{DeGennes, Kleinert} of length $l$---made of $N=l/a$ monomers of length $a$,  held under tension $f$  and subjected to a  torque $\Gamma$ by magnetic or optical tweezers~\cite{Neuman}, as in Fig~1. In a continuum limit, $\psi(s)$ represents the deviation from the straight filament configuration in the perpendicular plane and $s$ is the intrinsic coordinate. The   contribution from tension  to the energy density per unit length, neglecting subdominant  terms, is  $-f dh/ds=f \dot \psi^* \dot \psi/2=f (\dot \rho^2 +\rho^2 \omega^2)/2 $. Here  $dh$ is the experimentally measured change in distance $h$ between beads (Fig. 1).  
The simple-connectedness of the  space (and thus of the plane orthogonal to the experimental apparatus) is removed by considering another polymer, held straight, around which our polymer can tangle (Fig.~1).  Then the energy contribution to   the  torque is  $-\Gamma\int_0^l \omega ds$ as $\int_0^l \omega ds$ is the mutual angular deviation between the beads on which the torque acts. There are boundary conditions $\rho(l)=\rho(0)=\rho_0>\bar \rho$ at the extremes, but clearly not on the angular variable. 
Such a system is described by the energy in (\ref{H}) if  $\chi=f/T$, $\gamma=\Gamma/T$.  
One might also consider two identical polymers,  described by  $\phi_1$, $\phi_2$, and then $2\psi=\phi_1-\phi_2$ (the ``center of mass'' coordinate $\phi_1+\phi_2$ only contributing equipartition). In both cases $\bar \rho$ is the average of the two radii. 
This  problem of biased tangling  can  be extended beyond the experimental setting, for instance to the case of a floppy polymer tangling around another polymer with large persistence length (e.g. ssDNA tangling around helical  DNA~\cite{Mirkin}). 

 \begin{figure}[t]
\begin{center}
\includegraphics[width=.8\columnwidth]{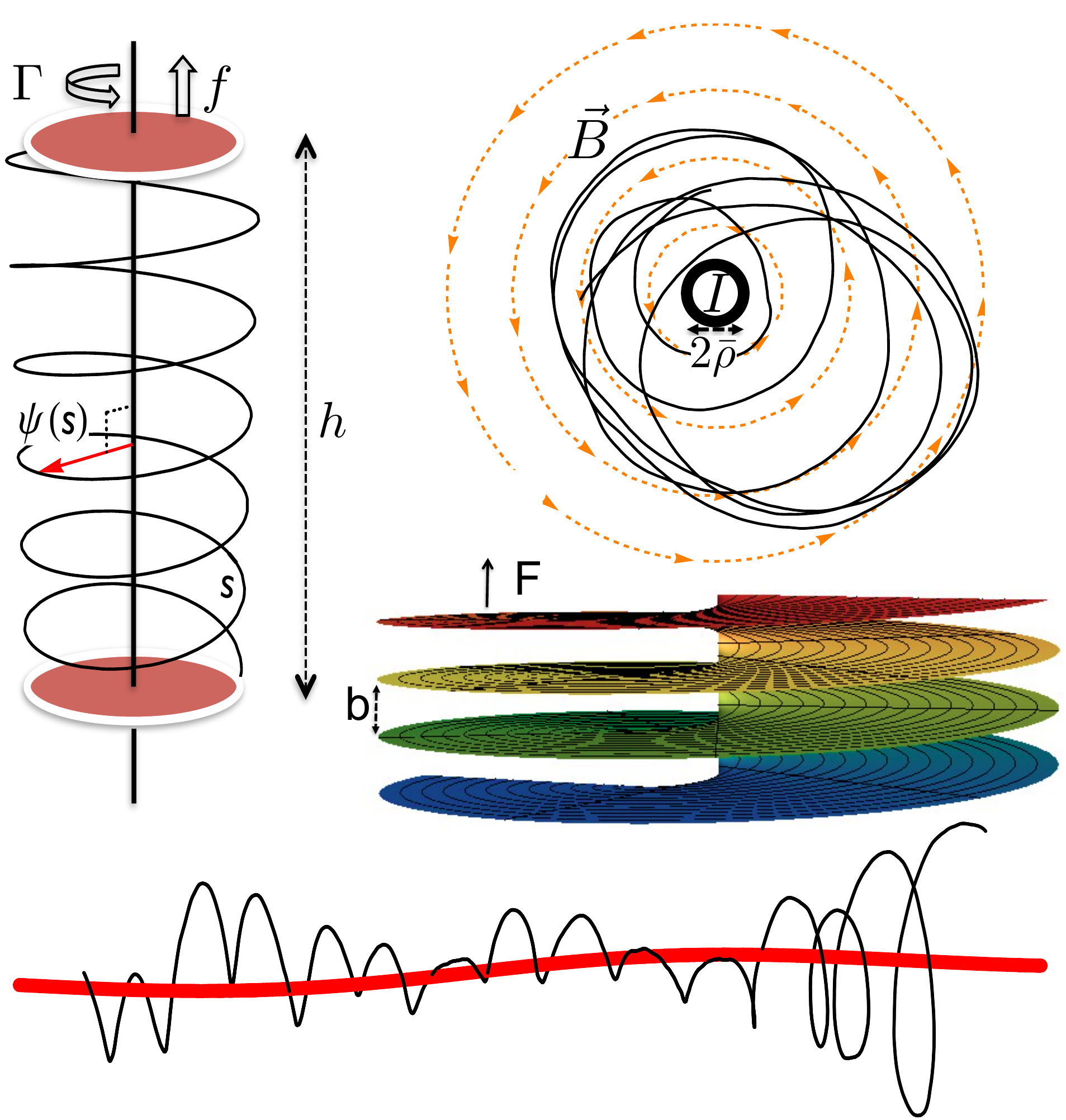}
\caption{Left: Schematics of a possible experimental setting; a fluctuating  polymer, here in a helical configuration, held at its ends with a tension $f$ between beads at distance $h$, subject to a torque $\Gamma$,  tangling around the straight polymer at the center,  and described by  the two-dimensional vector $\psi(s)$ (red arrow) in the plane perpendicular to $h$, while $s$ is the intrinsic coordinate.  Top right: a random chain with magnetized monomers of magnetic moment parallel to the tangent curling around a current $I$ generating a magnetic field $\vec B$. Middle right: the convection-diffusion-reaction (of dopant tracers) takes place on a Riemann surface (a screw dislocation) while the angular drift is provided by the applied field $F$ along the $z$ direction; $b$ is the Burger's vector. Bottom: a floppy polymer biased to tangle around a polymer of larger persistence length.}
\label{default}
\end{center}
\end{figure}

We now analyze the transition 
which  corresponds to the disappearance of the extensive winding of the two polymers. 
In the thermodynamic limit of large $l$,  (\ref{trace}) projects onto the lowest bound state of eigenvalue $\epsilon_0$---if there is one---giving  $\langle \omega\rangle= -T\partial_{\gamma} \epsilon_0\ne0$. 
The discrete spectrum of the operator in (\ref{Hs})  is known to disappear when $|\gamma|<1/2$, pointing to a transition at $T_c=2|\Gamma|$. For $T>T_c$ the contribution from the continuum spectrum in (\ref{trace}) is non-extensive in $l$,  and  the partition function effectively reduces to equipartition, or $Z\propto\chi^{-l/a}$, independent of $\gamma$, and thus  $\langle \omega \rangle=0$: above transition the extensive helical structure is lost~\cite{footnote1}.

When $T\rightarrow 2\Gamma^-$, the 1-D Schr\"odinger problem of (\ref{Hs}) on a half line with cutoff is well studied~\cite{Gupta}. Defining  $\nu^2=\gamma^2-1/4$,  the disappearing lowest bound eigenvalue can be approximated in the limit  $\nu\rightarrow 0^{+}$~\cite{Gupta}   as
\begin{equation}
\epsilon_0=-q e^{-\frac{2\pi}{\nu}}[1+O(\nu)],
\label{epsilon}
\end{equation}
with   $q^{-1}={ 3.17(2...)\bar \rho^2\chi}/T$. From (\ref{Omega}) and (\ref{epsilon})  it follows that the helical order parameter disappears at transition with infinite smoothness (Fig. 2)  as
\begin{equation}
\langle \omega\rangle\simeq {2\pi q } {\gamma \nu^{-3}}e^{-\frac{2\pi}{\nu}}\left[1+O(\nu^2)\right]
\label{omegac}
\end{equation}
and the transition is therefore of infinite order, as expected given its topological nature. 
\begin{figure}[t]
\begin{center}
\includegraphics[width=.85\columnwidth]{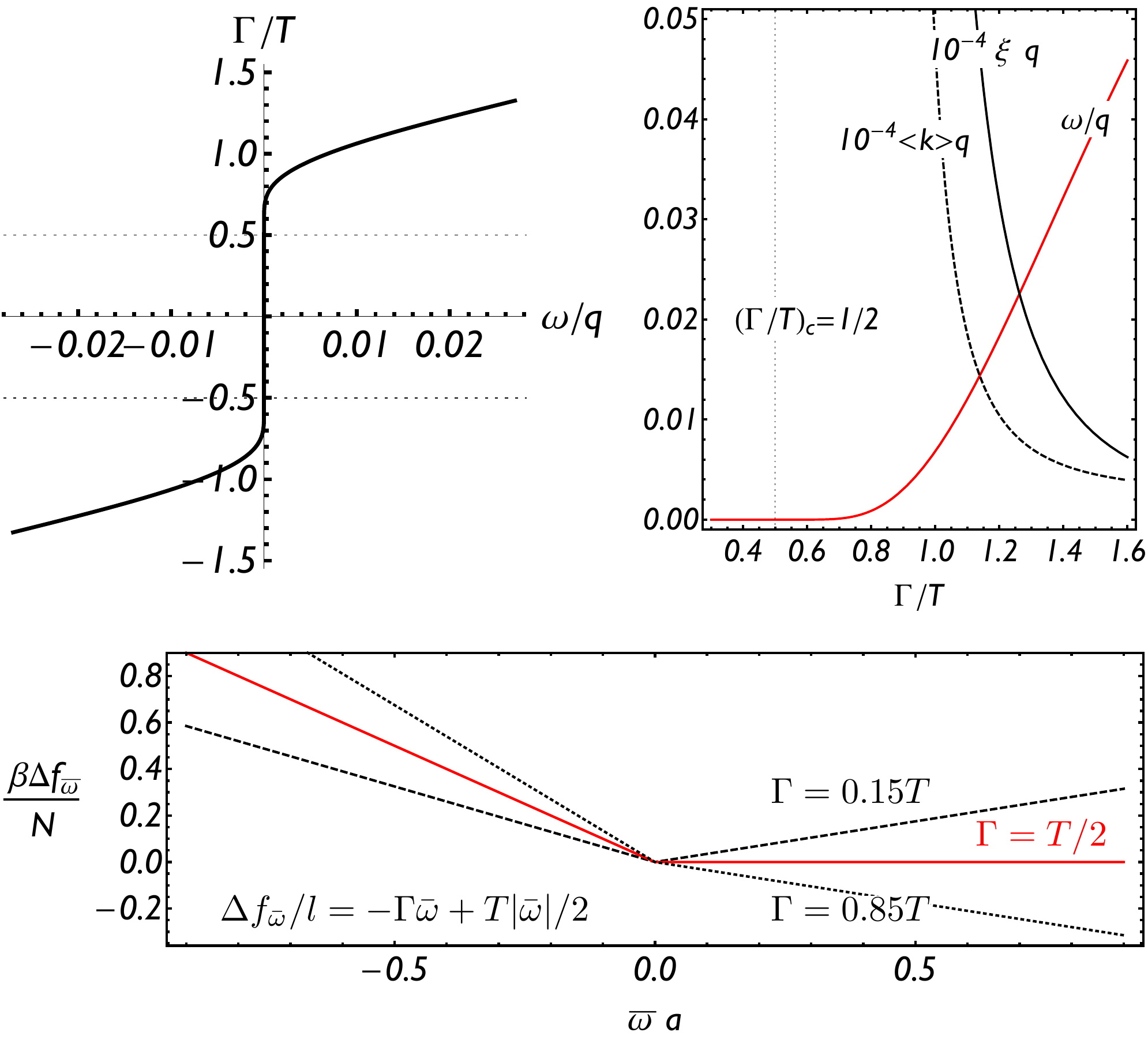}
\caption{Left: The transition in the $\gamma=\Gamma/T$ vs. $\omega$ plot; $|\Gamma/T|\to 1/2$ as $\omega\to 0$. Right: as $\Gamma/T \to 1/2+$ the average helicity $\langle \omega \rangle $ (red solid line) tends to zero exponentially fast, while both the generalized helical rigidity $\langle k\rangle \equiv 1/\langle k^{-1}\rangle$ (dashed black line) and the correlation length $\xi$ (black solid line) approach infinity with exponential behavior. Bottom: the   free energy density contributed by a definite helicity $\bar \omega$ from the heuristic argument in (\ref{heur}); for $\gamma<1/2$ we always have an increase of free energy, whereas at criticality ($\gamma=1/2$) all $\omega$  of the same sign of $\gamma$ are admitted; for $\gamma>1/2$ the largest possible winding $\omega$ (defined by  cutoff) provides the smallest lowering of the free energy.}
\label{default}
\end{center}
\end{figure}
The generalized rigidity $\gamma/\langle \omega \rangle$ approaches infinity exponentially fast at transition and therefore, from (\ref{kgen}), so does $\langle \rho^2 \rangle$. 

From  the statics of  (\ref{H}) we can gain some insight into the transition. 
The local minima of  ${\cal H}$ (for variations at fixed boundaries)   are uniform trajectories, or $\dot \psi(s)=0$, corresponding to a straight polymer parallel to the experimental axis (its statistical mechanics corresponds to planar oscillations).  However, for $\gamma\ne 0$, winding trajectories, which are not stationary points of the functional (in the sense that   $\delta{\cal H}/\delta \psi\ne0$), have lower energy. Among these, if $\rho$ is constrained, the global minimum is  a uniform helix with 
\begin{eqnarray}
\omega=k(\rho)^{-1}\gamma= \frac{\gamma}{\chi \rho^2},
\label{motion}
\end{eqnarray}
where $k(\rho)$ is the linear density of helical rigidity  {per unit length} and therefore $k^{-1}$  is the helical compliance introduced in (\ref{kgen}).  Note that the generalized  winding compliance of (\ref{kgen}) is simply the thermal average of the ``static''  compliance~(\ref{motion}). The energy of the helix is then 
\begin{equation}
\beta V(\rho)=-\frac{\gamma^2}{2\chi \rho^2}. 
\label{V}
\end{equation}
The cutoff provides a lower bound for the energy  at  $\rho=\bar \rho$  and helicity $\omega_0= k(\bar \rho)^{-1} \gamma$.

 The global minimum of the regularized  functional ${\cal H}$ in (\ref{H}) corresponds to a uniform winding around the pole, while its excitations are   straight polymers.  These two classes of trajectories are topologically distinct  in the non-simply-connected plane and  
a transition  can happen when their free energies   become degenerate. 
Indeed entropy reduction {is the cost of structure}: winding trajectories,  although  favored by energy, are  entropically disadvantageous compared to the non-winding ones: this competition   drives the transition and suggests the heuristic argument below. 

Summing over  fluctuations { in $\rho$ only}, while maintaining $\bar \omega$ fixed,  we obtain the partition function of a helix of uniform winding angle $\bar \omega$, or
%
 $Z_{\bar \omega}=e^{{ \gamma \bar \omega l}}\int [d\rho]\exp{\left[-\frac{ \chi}{2}\int_0^l  \left(\dot \rho ^2+\rho^2 \bar \omega^2\right)ds\right]}.$
%
The latter is  an harmonic  problem in $\rho$ when $\bar\omega\ne0$, while for  $\bar\omega=0$ it reduces to free oscillations and thus equipartition $Z_{\bar\omega}\propto \chi^{-l/a}$. For large $l$ we can project on  the lowest eigenvalue $|\bar \omega|/2$. 
By subtracting the  equipartition  free energy density  $(\ln \chi)/2a$ (obtained for $\bar\omega=0$) from the free energy density $f_{\bar \omega}=-T l^{-1}\ln Z_{\bar \omega}$, one arrives at
the (linear density of)  free energy difference  contributed by the helicity $\bar \omega$: 
\begin{equation}
\Delta f_{\bar \omega}/l=-\Gamma \bar \omega+ T|\bar \omega |/2.
\label{heur}
\end{equation}
Equation (\ref{heur}) implies that both energy ($-\Gamma \bar \omega$) and entropy ($\Delta s=-|\bar \omega|/2$) are reduced by a winding trajectory. As expected, their interplay drives the transition: for  $|\gamma|<1/2$, helical structures are suppressed, as any helicity  $\bar \omega \ne 0$ increases the free energy. However,  when $|\gamma|>1/2$,  the entropic cost of winding  can be offset by an energetic gain, and  helical structures of the same orientation of $\Gamma$ lower the free energy (Fig.~2)~\cite{footnote2}. As with the KT transition, the heuristic, entropic argument correctly predicts the critical temperature ($T_c=2|\Gamma|$).

Interestingly, the heuristic result in (\ref{heur}) is exact at transition with the substitution $\bar \omega \rightarrow \langle \omega \rangle$. In fact, from $f= T\epsilon_0$, $s=-\partial_Tf $, and (\ref{epsilon}), we  obtain  
\begin{equation}
\Delta s=-\frac{1}{2} \langle \omega \rangle\left[1+2\nu^2+ O(\nu^3)\right].
\end{equation}

Since the heuristic computation is based on a uniform winding angle, its exactness at transition suggests that the order parameter tends to uniformity at criticality.  As it disappears, its fluctuations must then  tend to zero, while their correlation length must approach infinity. The first statement is  proved true by differentiating  the expression for $\langle \omega\rangle$ in (\ref{omegac}) with respect to $\gamma$.  We establish the second one below.

Correlation lengths can be computed by introducing a varying external field $\gamma(s)=\gamma+\eta(s)$ with $\gamma$  uniform, as before. The winding correlation function
$G(s_1,s_2)=\langle \omega(s_1) \omega(s_2)\rangle-\langle \omega \rangle^2
$ %
is given by~\cite{Zinn-Justin}
\begin{equation}
G(s_1,s_2)=\frac{\delta^2\ln Z[\eta]}{\delta \eta(s_1) \delta \eta(s_2)}\Big |_{\eta=0}, 
\end{equation} 
where the new partition function $Z[\eta]$ 
is  still given by (\ref{trace}),  with the replacement $ l \hat H\rightarrow  l \hat H+ \gamma\int_0^l [\eta(s)/\chi\rho^2]ds$. 
  Standard perturbation calculations in imaginary time $s$ yield, for large $|s_1-s_2|$,

\begin{equation}
G(s_1,s_2)\propto e^{(\epsilon_0-\epsilon_1)|s_1-s_2|}.
\label{oo2}
\end{equation}
From (\ref{oo2}) and (\ref{omegac}) we have for the correlation length   
\begin{equation}
\xi=-(\epsilon_0-\epsilon_1)^{-1}\sim \exp \frac{2\pi}{\sqrt{\gamma^2-1/4}}
\label{xi}
\end{equation}
[since $\epsilon_1/\epsilon_0\sim \exp(-4\pi/\nu)$ at transition~\cite{Gupta}]. Equation ({\ref{xi}) is the same result of the KT model (but with $\Gamma$ replacing $T$). This can be expected as both transitions are topological in nature and both coincide with breaking of conformality ~\cite{Kaplan09}. 
However, unlike the KT case, an external field, $\Gamma$, breaks the symmetry of our problem and provides an order parameter, $\langle \omega \rangle$. Note that this symmetry breaking is absent in the quantum ISP problem (\ref{Hs}) before the Hubbard--Stratonovich transformation that leads to (\ref{H}).  

Our analysis also provides  a clear topological explanation for the well known  anomalous symmetry breaking of the ISP via renormalization~\cite{Gupta, Moroz10, Kaplan09, Essin}. 
The transition  corresponds to tangled fluctuations contributing to the partition function below 
 transition, and untangled above. These trajectories are topologically distinct in the punctured space. Taking the cutoff $\bar \rho\to0$ does not restore  simple-connectedness. Indeed the limit can be taken together with $\gamma\to (1/2)^+$ in such a way that $\langle \omega \rangle$ in (\ref{omegac}) remains finite, or  $\epsilon_0$ in (\ref{epsilon}), remain finite, which corresponds to the  quantum anomaly of the ISP in (\ref{Hs}).
   
Finally we show that the model corresponds to the theory for the Wilson line of the (non-dynamical) $U(1)$ gauge theory in $(0+1)$ dimensions for a field $\phi=\rho \exp i\alpha$:
\begin{equation}
\beta {\cal L_{\phi,A}}=\chi |(\partial_s-iA) \phi|^2/2+ \zeta A,
\label{gauge}
\end{equation}
which is invariant under $\phi\to \phi \exp i\Lambda$, $A\to A+\partial_s \Lambda$, if $\Lambda$ has periodic boundary conditions, $\Lambda(s)=\Lambda(0)$, and thus cannot  change the total winding number for $\phi$.  Then our previous coordinate $\psi$ can be considered as the Wilson line of the gauge theory (\ref{gauge}) in $\phi$ and $A$:
\begin{equation}
\psi(s)=\phi(s)e^{-i  \int^s \!A(t)dt}.
\label{paperino}
\end{equation}

From (\ref{paperino}), our winding parameter $\omega=\partial_s\alpha-A$ is the relevant coordinate, mapping a gauge-invariant functional manifold orthogonal to the gauge lines. Conversely,  the new coordinate $\eta=\partial_s\alpha+A$ describes the gauge trajectories: 
 indeed $\eta$ flows with the gauge as $2\Lambda$. 

With this in mind (\ref{gauge}) can be rewritten as
\begin{equation}
\beta {\cal L_{\phi,A}}=\chi \left(\dot \rho ^2+\rho^2 \omega^2\right)/2-\zeta (\omega-\eta)/2,
\label{pippo}
\end{equation}
since $2A=\eta-\omega$. In the partition function the term $\zeta \eta/2$ factors into the irrelevant integration over  the gauge trajectories (while  the Faddeev-Popov determinant is inconsequential, the theory being abelian), and we are effectively left with $\beta {\cal L_{\phi,A}}=\beta {\cal H}_{\psi}$, our energy for $\psi$ given by ({\ref{H}), {\it but with $\gamma=\zeta/2$}. 

We see now that the transition in the $U(1)$ gauge theory corresponds to $\zeta=1$, for which the expression $\exp( i\zeta \int^s\!\!\!Adt)$   becomes invariant toward a gauge with {\it free boundaries} in the thermodynamic limit. Indeed  the allowed values $\Lambda(s)-\Lambda(0)= 2\pi n$ correspond to the change of winding number per unit length $2\pi n/l$ which approaches continuum as $l\to \infty$. At transition the gauge space extends  to transformations that can change the average winding number of the field: that is natural, as the two functional spaces (winding and unwinding trajectories), are only topologically distinct at fixed boundaries, a constraint removed at transition. 

Before concluding, we propose other realizations. Considering  $\psi$ as a two-dimensional vector $\vec x$ we write (\ref{H}) as 
 \begin{equation}
 \beta{\cal H}=\chi |\dot{\vec x}|^2/2-\gamma \dot{\vec x}\cdot \vec{w}(x),
 \label{H1}
 \end{equation}
where $\vec w=\vec \nabla\theta(\vec x)= {\hat e_3 \wedge \hat x}/{|\vec x|}$ is the field of an elementary vortex ($\hat e_3$ is the unit vector perpendicular to the plane, $\theta(\vec x)$ the angular coordinate of $\vec x$). 
 If $\gamma \vec w$   represents a magnetic field generated by a current $I$ perpendicular to the plane,  the path integral in (\ref{Zeta})  describes the probability distribution for a magnetized ideal random chain~\cite{Kleinert} in the magnetic field, where each monomer has a magnetic moment $\vec m \propto d\vec x$ and $\gamma=\mu_0 |m| I/2\pi T$ ($\chi=2/a$). 
 
Finally,  if $l \to t$, $\chi=1/2D$, where $D$ is the diffusivity,  and if $\{(\vec x_0,t_0)\to(\vec x, t)\}$ are chosen as boundary conditions for the trajectories in the path integral, then, from (\ref{Zeta}) and (\ref{H1}),  $P(\vec x,t; \vec x_0,t_0)\propto Z$ describes the solution of the following convection-diffusion-reaction equation
\begin{equation}
D^{-1}\dot P= \vec \nabla \cdot\left(\vec\nabla P-\gamma \vec w P\right) +P{\gamma^2} w^2 /4
\label{diffusion}
\end{equation}
on a helical Riemann surface under drift $D\gamma \vec w$. The Riemann surface can be a screw dislocation in a material where only in-plane diffusion is allowed~\cite{Inomata}. Then drift can come from a field $F\hat e_3$  parallel to the Burger's vector $\vec b=\hat e_3 b$ (Fig.~1), since 
$z/b= \theta/2\pi$. Then  $\gamma=2\mu F b/D$, where $\mu$ is the mobility ($2 \mu F b$ is the vorticity of the drift): the transition corresponds to a competition between diffusivity and drift. A non-zero order parameter in (\ref{Omega}) implies a uniform (in time) climbing of the dislocation, or {$z\sim h\theta/2\pi\sim h{\langle\omega \rangle}t/2\pi$}.

In conclusion we have reported a topological winding/unwinding transition connected with the quantum loss of conformality of the attractive ISP. The quantum anomaly of the potential at strong couplings is related to the non-simple-connectedness of the manifold that allows for topological distinction.  Below transition winding topologies are energetically favored, although entropically unfavored, and vice versa above transition. We have proposed possible physical applications including  polymer physics and  diffusion-convection-reaction. In particular, an experiment in single molecule manipulation of an appropriate floppy polymer (Fig.~1) could reveal the transition: at room temperature the critical torque is   $\sim$2 pN$\times$nm.

C. Nisoli is grateful to P. Lammert  for discussions. This work was carried out under the auspices of the National Nuclear Security Administration of the U.S. Department of Energy at Los Alamos National Laboratory under Contract No.~DEAC52-06NA25396.


\begin{thebibliography}{}
\bibitem{footnote2} The fact that $\Delta f_{\bar \omega}$ becomes unbounded from below in $\bar \omega$ when $T>2|\Gamma|$ is a consequence of neglecting the cutoff  in this heuristic computation.
\end{thebibliography}
\end{document}